# VIPER – Student research on extraterrestrial ice penetration technology

F. Baader, M. Reiswich, M. Bartsch, D. Keller, E. Tiede, G. Keck, A. Demircian, M. Friedrich, B. Dachwald
Department of Aerospace Engineering
FH Aachen University of Applied Sciences
Aachen, Germany

K. Schüller, R. Lehmann, R. Chojetzki, C. Durand, L. Rapp, J. Kowalski
RWTH Aachen University
Aachen, Germany

R. Förstner
Institute of Space Technology and Space Applications
Bundeswehr University
Munich, Germany

*Abstract*—Recent analysis of scientific data from Cassini and earth-based observations gave evidence for a global ocean under a surrounding solid ice shell on Saturn's moon Enceladus. Images of Enceladus' South Pole showed several fissures in the ice shell with plumes constantly exhausting frozen water particles, building up the E-Ring, one of the outer rings of Saturn. In this southern region of Enceladus, the ice shell is considered to be as thin as 2 km, about an order of magnitude thinner than on the rest of the moon. Under the ice shell, there is a global ocean consisting of liquid water. Scientists are discussing different approaches the possibilities of taking samples of water, i.e. by melting through the ice using a melting probe. FH Aachen UAS developed a prototype of maneuverable melting probe which can navigate through the ice that has already been tested successfully in a terrestrial environment. This means no atmosphere and or ambient pressure, low ice temperatures of around 100 to 150K (near the South Pole) and a very low gravity of 0,114 m/s^2 or 1100 µg. Two of these influencing measures are about to be investigated at FH Aachen UAS in 2017, low ice temperature and low ambient pressure below the triple point of water. Low gravity cannot be easily simulated inside a large experiment chamber, though. Numerical simulations of the melting process at RWTH Aachen however are showing a gravity dependence of melting behavior. Considering this aspect, VIPER provides a link between large-scale experimental simulations at FH Aachen UAS and numerical simulations at RWTH Aachen. To analyze the melting process, about 90 seconds of experiment time in reduced gravity and low ambient pressure is provided by the REXUS rocket. In this time frame, the melting speed and contact force between ice and probes are measured, as well as heating power and a two-dimensional array of ice temperatures. Additionally, visual and infrared cameras are used to observe the melting process.

## I. Introduction

The Vaporizing Ice Penetration Experiment on a Rocket (VIPER) is a REXUS-Experiment to collect scientific data while melting probes penetrate samples of water ice in low temperature, low gravity and low ambient pressure on a sounding rocket. It is managed and realized as a student project at FH Aachen University of Applied Sciences (FH Aachen UAS). In total, around 15 students from FH Aachen UAS and RWTH Aachen University are working voluntarily at VIPER since November 2016.

VIPER is funded by FH Aachen UAS and by the REXUS/BEXUS programme. Additional travel budget is provided by the Hans Hermann Voss Foundation.

### A. Student Projects at FH Aachen UAS

In many courses at FH Aachen UAS, students are encouraged to establish student projects on a voluntary basis. These are usually related to the student's field of study, but not part of the official curriculum. Based on their own ideas, students can submit project proposals to a panel of professors in a competitive manner once a year. The best proposals qualify for monetary funding including material purchase and travel expenses. Also non-monetary support will be provided in the form of working space, possibility of using laboratories and test facilities at the university and support by a supervising professor. Projects are not limited to university funding, but may also apply for external support.

The student projects are managed and conducted totally by the students themselves. The aim of the university is to set up an environment, where students can apply their theoretically acquired skills to real problems in an interdisciplinary way. Students learn to estimate their skills and personal strengths realistically and to take responsibility for the budget and a team.

All student work is conducted on a purely voluntary basis and carried by the team's motivation to reach a common goal.

As in 2018, there are more than 25 projects currently running. Examples from different faculties are short-movie productions, new city planning approaches, a formula student racing team, development and testing of unconventional aircraft designs or design of planetary exploration equipment.

*B. Background & Scientific Objectives*

The presence of subglacial liquid water on the icy moons of our Solar System [1] implies the possibility of environmental conditions suitable for life. Especially Enceladus, the cryo-volcanically active moon of Saturn, seems to be a promising candidate and there is some hope in the scientific community that an Enceladus exploration mission might unravel the existence of extra-terrestrial life. Next generation mission concepts consider orbiting and sample-returning of plume material [2, 3]. If these further strengthen any evidence for life, a mission that samples and analyses the subglacial liquid directly would be the natural next step [3, 4]. In order to access the extra-terrestrial subglacial water reservoirs, a thick ice layer must be penetrated.

A very promising approach for this challenging task is to use a thermal melting probe [4]. Melting probes enforce ice penetration by heating, such that the ice in the vicinity of the probe melts. In combination with an applied force, e.g. given by the probe's weight, motion through the ice is possible. The amount of necessary power to heat the probe roughly scales with the cross-sectional area of the melting channel. Therefore, a melting probe typically looks like an elongated cylinder with a heated melting head. In comparison to other ice penetration technologies, e.g. hot water or mechanical ice drilling, its design is smaller, lighter, and comprises less complex mechanical devices.

Melting probes are not a novel technology as they have been already successfully applied for terrestrial research in the 1960's [5, 6]. Although the basic concept remained the same, over the recent years more advanced designs have been reported and tested [7, 8, 9, 10]. More complex technical designs have also necessitated the development of advanced mathematical models [11, 12, 13, 14].

For the majority of potential target environments for extra-terrestrial melting probes, the ambient pressure is below 6.1 mbar, which is approximately the triple point of water. Hence, the ice will sublimate during the initial phase of ice penetration. The produced vapour will eventually refreeze at the probe's hull or at other critical locations. Under certain circumstances, this can cause stall of the probe.

No melting probe has yet been used in an extra-terrestrial environment, in which low temperatures, a low environmental pressure as well as a small gravity is present. Towards an extra-terrestrial melting probe mission, a lot of testing and modelling effort is required. Two of these extreme conditions can be tested on the ground in a thermal-vacuum chamber, as it has been done in [15]. However, the combination of all three environmental conditions has not been studied so far. From experiments, it is known that both, a low pressure and a low temperature have a dramatic impact on the melting velocity of a melting probe. In order to better understand the role of gravity regarding ice penetration by melting probe technology, we developed VIPER.

*C. The REXUS/BEXUS programme*

Based on a bilateral agreement between the German Aerospace Center (DLR) and the Swedish National Space Board (SNSB), REXUS/BEXUS (Rocket resp. Balloon Experiments for University Students) allows students to have their scientific experiments or technology demonstration to be launched on a sounding rocket or a high-altitude balloon from Esrange Space Center in Sweden. Each year, student payloads for two rockets and two balloons are selected. Usually around half of the available space is given to experiments from Germany, while the other half is open to experiments from all member states of the European Space Agency (ESA).

Successful applicants are granted a limited sponsoring for material and travel expenses and workshops introducing standards and working methods for space application.

FH Aachen already participated in the REXUS programme twice to qualify a vibration isolation mechanism for sounding rocket experiments (VIBRADAMP and ADIOS) [16] and in-flight modal analysis of the rocket structure [17].

*D. VIPER: Experiment Overview*

Utilizing the reduced-gravity phase in flight, VIPER is going to simulate penetration of an icy moon with low gravity and nearly no atmosphere. To do so, it carries three separated ice samples and three independently deployable heated melting probes. The melting probes are securely locked during launch and descent of the rocket to prevent damage to the experiment or the rocket itself during high-load flight phases. While approaching the reduced gravity phase of the flight, the melting probes become unlocked and three springs (one per probe) are pushing them into the ice samples.

Measured data while proceeding through the ice include penetration depth, differential pressure inside the experiment container relative to the rocket's ambient pressure and a temperature field with high spatial resolution for each of the ice samples. Additionally, two cameras (one for visible light and one for infrared footage) are documenting the experiment.

The VIPER experiment is located inside a standard REXUS rocket-module with a length of 300 mm and a diameter of 356 mm. Experiment space is divided into two zones, the wet zone and the dry zone. The dry zone is connected to the rest of the rocket's interior. It contains most of the electronics components, such as the experiment's communication and control systems. Furthermore, it contains data processing and storage components and a power management system including additional batteries. Water (i.e. ice or vapour respectively) is only allowed inside the wet zone. Here, the ice samples are located as well as melting probes, locking mechanism, cameras and most of the sensors. While penetrating the ice, sublimation produces a steady vapour mass flow, which can leave the wet zone through four venting holes connected symmetrically to the outer rocket structure by tubes. This guarantees to keep the experiment's inner pressure below the triple point of water, which is necessary to simulate the conditions on an icy moon.

The experiment data and camera footage will be stored locally on redundant flash storage. Additionally, most of the data and camera footage will be transferred to the ground station via a downlink.

## II. SPACE EDUCATION

### A. Provided Infrastructure at FH Aachen

FH Aachen UAS offers Bachelor's and Master's degree courses in aerospace engineering at a dedicated faculty. Multiple laboratories for automotive and aerospace research provide useful tools to design, manufacture, qualify and test space equipment. Talking about space application, these include a manufacturing workshop, rapid prototyping facility, electronics lab, thermal imaging systems, vacuum chambers in different sizes, thermal simulation chambers and vibration simulation facilities. During the development of VIPER, some of these tools were heavily utilized (see chapter IV).

VIPER was provided their own office and dedicated working space in the laboratory, which are in their own responsibility.

### B. Experiencing research facilities and the science community

In the eyes of many students, getting access to laboratories and research facilities symbolizes the next-higher step between theoretical studies and real research and engineering work. Closing this gap is seen essential not only to sharpen the student's sense of responsibility, but the feeling of appreciation is also valuable in motivating them to graduate successfully in a demanding course like aerospace or electrical engineering. In addition, the practical aspect of designing and manufacturing parts is a very useful extension to deepen and apply the theoretical background from the lectures. Students learn that the best theoretical option is not always the best regarding costs, practicability or availability of different components.

Visiting and working at facilities and laboratories of ESA and DLR gives the students a way better idea of what designing and manufacturing a rocket experiments and space hardware in general is about than any lecture at university or short excursions can do. Apart from that, the REXUS programme offers the possibility to work together with highly experienced space engineers and scientists. Also among the teams themselves, an informal network between the participants all across Europe formed after experiencing similar problems since being selected to participate.

In the opposite direction, students used VIPER in schools, to get pupils interested in space related technology and STEM in general. The team presented its experiment and the REXUS/BEXUS programme at schools in Aachen and Berlin, to give an insight into high altitude and space research while still at school.

### C. A critical view on professional standards in student projects

Nowadays, space industry is predominantly working according to common standards by the European Cooperation for Space Standardization (ECSS). For companies, ECSS standards are providing a frame for quality management, as well as project management methods. Trying to transfer them as a whole to small student projects like VIPER will probably lead to perfect documentation on one hand, but frustrated students and no hardware at all on the other hand. This can even be extended to university research projects. ECSS standards should not be put aside as useless or even obstructive, though. Interpreted as guidelines how it should be done, they became apparent to be a useful tool in the VIPER project.

Taking project management as an example, it was a demanding challenge to find the best balance between following accurately elaborated work packages and letting people follow their own idea of what to do next. Mostly, the second is more productive at first, since working becomes more dynamic and flexible. But especially when this work is affecting interfaces, problem arise on a regular basis. Students learn to understand there is more than one side of the coin, e.g. defining an interface two weeks later may be easier for one person, but detains another person depending on that interface from effectively proceed with his working package. Generating a level understanding for management decisions can be essential for students, whose eventual jobs will be in a highly managed environment, such as the space industry.

A similar approach may be useful for actual engineering, e.g. in electronics and PCB-design: Standards may define a minimum distance between conducting lines for a certain current. Taking this as a general design rule is totally reasonable and advisable. But if in a later design iteration changes are necessary and one has to decide between a strikingly increase in either budget, effort or complexity on one side and a slight undercut of the minimum distance on the other side, the second solution may be worth consideration.

Some common practice from industry can give real advantages to small space engineering projects, though. Best example may be well-conceived requirements for all aspects of the application or experiment, as well as a sound risk analysis. Both makes students think about their project in every detail. In contrast to working with just a rough picture in mind, this method ensures that all team members are working towards a common goal with a precise sense of how this goal looks like. Students will learn, that in this way, a solid base of requirements helps preventing miscommunication inside the team.

Reviews (conducted by an external or internal review board likewise) on a regular basis became apparent to be a powerful tool not only to report progress, but also making the team aware of weaknesses in their current design. Additionally, students get used to confidently present their work in front of a critical group of experts, which may become useful later on in presenting research on scientific conferences.

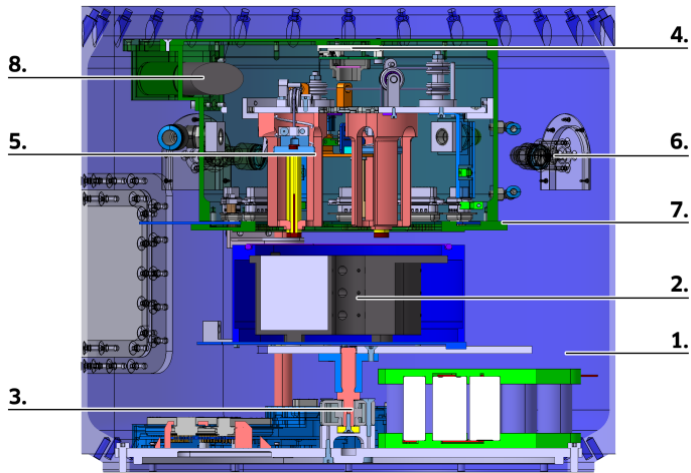

Fig. 1: VIPER experiment design

## III. EXPERIMENT DESIGN

The VIPER experiment design was developed primarily to fulfil the given scientific requirements. In addition, since the experiment shall fly on a rocket, there where requirements given by the launch provider. Most prominent to mention is that water contamination of any other inner part of the rocket than our own module has to be avoided in a reliable way. Components labelled in Fig. 1 are described in the following. Including the rocket module (1.), the total mass of the experiment is 13.7 kg.

### A. Mechanical Design

The Ice Sample Container Assembly (ISCA, 2.) represents the center of the whole experiment. It assembles three cylindrical, triangular arranged containers, which store solid water ice samples. In each ice sample, nine PT100 temperature sensors are integrated to measure a three-dimensional temperature field while the melting probe penetrates the ice. The ISCA is inserted shortly before launch and lifted by a combination of a self-locking worm gear and lead screw (3.) until locking up against the Cupola.

The Cupola (7.) is the sealing counterpart on top of the ISCA. It is located and fixed in the upper section of the rocket module and contains the Heat Probe Pushing Mechanism (HPPM, 5.), cameras (one optical and one IR camera, 4.) as well as a venting system (6.) which is connected to the rocket's hull. Two feedthrough PCBs on top of the bottom side of the Cupola provide electrical connection with D-SUB connectors from the outside to the inside. Because water (or vapour) can leave the Cupola only through venting tubes, there is no way for it leaking into the rest of the rocket module.

One HPPM is placed above each ice sample. Two springs with spring rate R=388 N/m (max. 32N) and one with R=867 N/m (max. 78N) push the melting probes simultaneously against the ice samples during the experiment. The melting probes are relieved by a self-locking linear spindle drive, powered by a brushless direct current (BLDC) motor (8.), providing a maximum melting distance of 45mm. The force of the linear drive to relieve or to retract the melting probes is transmitted by a highly flexible stainless-steel wire. A special arrangement of guide pulleys ensures an optimal use of space. The spindle drive provides a maximum force of 189 N, while the maximum force of the three springs equals 142 N in the fully retracted state. At the carrier of the spindle drive, a set of tension springs is used to keep the steel cables tensioned. The melting probes contain heating cartridges inside a copper shell. Two of them are powered with 70W and one with 35W. Except for the tip, the copper shell is additionally covered with PTFE insulation. Using an optical encoder for each probe, the melting distance, speed and contact force may be derived.

### B. Electronic Design

The electronic system divides into three major subsystems: data acquisition from the cameras and other sensors, actuator (motor and melting probe control) control and power management.

On the mainboard, a STM32 microcontroller is located along with two redundant flash chips for sensor data storage, and all sensors are connected to the mainboard.

The two cameras, a white/black-CMOS camera from iDS and a Flir Lepton III LWIR-camera, are connected via USB, SPI and I2C to a Raspberry Pi Compute Module 3 mounted on a custom board. This appeared to be necessary for size reasons and because the USB-Hub normally used on Raspberry Pis is not capable of the data rate generated by the camera.

Power management: A custom powerboard converts 28V/1A supplied by the service module to 5V and 3.3V and distributes the power to all electronic components except the melting probes, which are supplied by our own battery through custom 24V boost converters. To supply three melting probes with in total 175W, different battery types were evaluated. Where the best trade-off appeared to be using a pack made of 9 cells COTS high-power-type NiMH from Panasonic [BK300SCP]. Battery packs were qualified by ourself for usage at high currents and in vacuum. Since batteries provide a voltage of between 8V and 12V during discharge, but the heating cartridges require a voltage of 24V, boost converter are required. Thermal tests inside the vacuum chamber showed, that overheating issues may occur with the first boost converter revision using normal FR4-PCBs. Therefore, boost converters were optimized to have a single side layout to be used on an Insulated Metal Substrate PCB to ensure sufficient cooling.

In total, eleven printed circuit boards were designed, some of which are noticeable. Two PCBs only have connectors and wires mounted and are designated to provide a proper sealed feedthrough for connections into the wet part of the experiment. Four PCBs (mainboard, powerboard, pi-board and the boost-converter) are quite complex. We nearly exclusively used SMT components on our board because they are small, less fragile to vibrations and faster and easier to assemble. To keep the design process simple, it was aimed for double-sided PCBs with a single assembly side, although this last was not possible for mainboard and powerboard. All huge components are fixed by screws or are glued onto the boards.

## C. Software Design

For REXUS experiments there are basically three software-parts to consider. First, all generated data shall be recorded, processed and filtered. Also the experiment has to be controlled, in our case melting probes are released and later retracted by a BLDC motor, the probes are switched and sensors and are turned on. These actions have hard real-time requirements and therefore, software runs on a microcontroller without operating system.

Second: The (so-called) groundstation receives data during flight, processes, filters and displays it to the groundstation operator. It has to display experiment data clearly and has to provide protection against accidental misuse. Furthermore, commands may be sent from groundstation to experiment module to enable a test mode for cold tests.

Third part is the communication between groundstation and experiment module, which is also used inside the experiment to communicate between the two processors. The REXUS rocket provides a serial connection of 34800 baud, bidirectionally transmitted via cables on ground, respectively only from rocket to ground using a wireless connection after launch. Since connections might be interfered, resulting in data loss or bit flips, an error detection was implemented. The code for the communication is generated by a self-written code-generator and shared between all platforms to minimize the possibility of errors and to provide the maximum test-coverage.

For all software parts, C++ was chosen as programming language. C++ as a compiled language has the advantage of showing errors during compilation and earlier than any interpreted language, very important for maximum reliability. Where possible, already existing and tested frameworks were used. Code for the groundstation is written with Qt. The microcontroller is programmed using the xpcc framework [http://xpcc.io/] that provides cooperative multi-tasking and many drivers for sensors and actuators used in our experiments. Additionally, we wrote some drivers that did not exist before and upstreamed them to the framework.

## D. Thermal Design

VIPER consists of a purely passive thermal control system involving proper materials, phase change materials and insulation.

The timespan between ISCA insertion and rocket launch is in the order of 50 minutes or even larger. In order to meet the requirement of ice sample temperatures smaller than -30 °C before the ice penetration starts, the ISCA is filled with self-produced dry ice snow. The resulting low temperatures induce condensation while the rocket is on the ground. The ISCA is therefore utilized with water-absorbing material on its circumference.

The three melting probes, on the other hand, induce high thermal loads at a short timespan to the HPPM assembly. We therefore use custom 100 W heating cartridges of 50 mm length with an inhomogeneous power distribution (99 % of the total power is located at the first 20 mm near the tip). It should be noted that we operate the three heating cartridges at 70 W, 70 W and 35 W, respectively. Both, numerical thermal analyses as well as validation experiments revealed that using these heating cartridges reduces the maximum temperature of the whole assembly drastically to an acceptable value of approximately 55 °C.

Additional numerical analyses have been conducted in order to calculate the minimum and maximum temperatures at critical locations including the thermal loads due to rocket launch, e.g. at the bulkhead where the boost converters are located. The results show that the temperature-critical parts of VIPER will sustain all expected thermal loads during launch and operation.

## IV. ENGINEERING APPROACHES, STICKING POINTS AND LESSONS LEARNED

A well-known engineering principle for applications appearing outstanding complex is the KISS approach ("Keep is simple and stupid"). Slightly modified to "Keep it safe and simple", KISS turned out to be extremely valuable for a project like VIPER with its limited budget, workforce and experience, while handling strict requirements concerning space, weight, autonomy, and thermal management.

For structural design, students used the university's CAD tool (CATIA V5-6R2012) to generate a detailed three-dimensional model of the whole experiment. With respect to the very limited space available inside the rocket, CAD proved useful for defining hardware interfaces, checking accessibility and estimating masses. Initially, it was planned to make use of rapid prototyping technology to manufacture sensor mounts casings and similar components which are not exposed to high mechanical or thermal loads. While we did not experience issues deploying 3D-printed parts into vacuum, the final experiment design nevertheless contained only a few minor components manufactured this way. Though manufacturing effort and costs were extremely low compared to traditional milling lathing from aluminium pre-product, but this advantage was totally drained by manual postprocessing to achieve required tolerances and geometries. This was true especially for more complicated components, such as electronics casings. Several components, for which additive manufacturing seemed beneficial at first were later on redesigned and manufactured again using classic technologies.

To conduct thermal analyses on specific items, detailed CAD-models were simplified again to reduce computational effort in ANSYS Workbench to a reasonable level. After applying correct boundary conditions, thermal simulations were confirmed by experimental pre-tests with adequate accuracy.

Essential for lowering the cost and design effort is the usage of commercial off-the-shelf (COTS) components, especially talking about electronics design. Still, a certain mental flexibility should be maintained when changes in requirements arise. Sticking to a COTS solution too vigorously may lead to a huge increase in effort or insufficient system performance. In example for VIPER, using COTS subassemblies in the electronics subsystem was constrained by requirements especially for size, weight and precision. Most of the power electronics and communication setup had to be developed on custom printed circuit boards (PCBs). Identifying the point when to switch to a custom solution is vital for taking advantages from COTS components, it requires some engineering experience, though.

While designing the experiment, a strongly interdisciplinary working attitude of all team members seems very advantageous to solve problems. One great example from the VIPER design process is the electrical feed-through of nearly 200 lines connecting sensors, melting probes and motors in the wet zone with electronics in the dry zone. This feed through had to be vacuum and vapour sealed, but there was also only very small installation space on a curved surface available. For COTS components this emerged to be a serious problem, which had to be solved. Finally, only the idea of using a PCB as lower ISCA cover and cable feed through in one part made it possible to have the required amount of sensors installed in the wet zone. Sensors and power lines are soldered to the PCB, which then is glued below the ISCA in a second step. Following this, standardized D-Sub High Density Connectors provide a pluggable interface to electronic parts in the dry zone.

## V. Conclusion

Student projects provide an opportunity to educate students beyond their courses' curricula and to help them gaining work experience before graduation. In case of the VIPER team, 15 students designed a full rocket experiment to be launched on the REXUS 23 sounding rocket. VIPER is going to contribute to research related to the exploration of icy moons by investigating the melting performance of melting probes in low pressure and low gravity regimes. The data will be compared to similar experiments on Earth. The results will help to develop and validate computational models for melting probes in extraterrestrial environments.